\documentclass[twocolumn, prx, superscriptaddress,notitlepage]{revtex4-2}
\usepackage{graphicx}
\usepackage{dcolumn}
\usepackage{bm}
\usepackage[usenames,dvipsnames]{color}
\usepackage[most]{tcolorbox}
\usepackage{multirow}
\usepackage{gensymb}
\usepackage[normalem]{ulem}
\usepackage{CJK}
\usepackage{comment}
\usepackage[colorlinks, linkcolor=blue,anchorcolor=blue,citecolor=blue,urlcolor=blue]{hyperref}
\usepackage{amssymb}
\usepackage{pifont}
\usepackage{physics}
\usepackage{natbib}

\usepackage{booktabs}
\usepackage{colortbl}

\usepackage{color,soul}
\usepackage[mathscr]{euscript}
\begin{document}
\begin{CJK*}{UTF8}{}
\title{Blind Quantum Machine Learning with Quantum Bipartite Correlator}



\author{Changhao Li}
\email{changhao.li@jpmchase.com}
\affiliation{
  Global Technology Applied Research, JPMorgan Chase, New York, NY 10017 USA}
\affiliation{
  Research Laboratory of Electronics, Massachusetts Institute of Technology, Cambridge, MA 02139, USA}
\affiliation{
  Department of Nuclear Science and Engineering, Massachusetts Institute of Technology, Cambridge, MA 02139, USA}

\author{Boning Li}
\affiliation{
  Research Laboratory of Electronics, Massachusetts Institute of Technology, Cambridge, MA 02139, USA}
\affiliation{
  Department of Physics, Massachusetts Institute of Technology, Cambridge, MA 02139, USA}

\author{Omar Amer}
\affiliation{
  Global Technology Applied Research, JPMorgan Chase, New York, NY 10017 USA}

\author{Ruslan Shaydulin}
\affiliation{
  Global Technology Applied Research, JPMorgan Chase, New York, NY 10017 USA}

\author{Shouvanik~Chakrabarti}
\affiliation{
  Global Technology Applied Research, JPMorgan Chase, New York, NY 10017 USA}
  
\author{Guoqing Wang}
\affiliation{
  Research Laboratory of Electronics, Massachusetts Institute of Technology, Cambridge, MA 02139, USA}
\affiliation{
  Department of Nuclear Science and Engineering, Massachusetts Institute of Technology, Cambridge, MA 02139, USA}

\author{Haowei Xu}
\affiliation{
  Department of Nuclear Science and Engineering, Massachusetts Institute of Technology, Cambridge, MA 02139, USA}

\author{Hao Tang}
\affiliation{Department of Materials Science and Engineering, Massachusetts Institute of Technology, Cambridge, MA 02139, USA}

\author{Isidor Schoch}
\affiliation{
  Department of Nuclear Science and Engineering, Massachusetts Institute of Technology, Cambridge, MA 02139, USA}

\author{Niraj~Kumar}
\affiliation{
  Global Technology Applied Research, JPMorgan Chase, New York, NY 10017 USA}

\author{Charles Lim}
\affiliation{
  Global Technology Applied Research, JPMorgan Chase, New York, NY 10017 USA}


\author{Ju Li}
\email{liju@mit.edu}
\affiliation{
  Department of Nuclear Science and Engineering, Massachusetts Institute of Technology, Cambridge, MA 02139, USA} 
\affiliation{Department of Materials Science and Engineering, Massachusetts Institute of Technology, Cambridge, MA 02139, USA}
 
\author{Paola Cappellaro}  
\email{pcappell@mit.edu}
\affiliation{
  Research Laboratory of Electronics, Massachusetts Institute of Technology, Cambridge, MA 02139, USA}
\affiliation{
  Department of Nuclear Science and Engineering, Massachusetts Institute of Technology, Cambridge, MA 02139, USA}
\affiliation{
  Department of Physics, Massachusetts Institute of Technology, Cambridge, MA 02139, USA}

\author{Marco Pistoia} 
\email{marco.pistoia@jpmchase.com}
\affiliation{
  Global Technology Applied Research, JPMorgan Chase, New York, NY 10017 USA}

\begin{abstract}
Distributed quantum computing is a promising computational paradigm for performing computations that are 
beyond the reach of individual quantum devices. Privacy in distributed quantum computing is critical for maintaining confidentiality and protecting the data 
in the presence of untrusted computing nodes. In this work, we introduce novel blind quantum machine learning protocols based on the quantum bipartite correlator algorithm. Our protocols 
have reduced communication overhead 
while preserving the privacy of data from untrusted parties. 
We introduce robust algorithm-specific privacy-preserving mechanisms with low computational overhead
that do not require complex cryptographic techniques.
We then validate the effectiveness of the proposed protocols through complexity and privacy analysis. Our findings pave the way for advancements in distributed quantum computing, opening up new possibilities for privacy-aware machine learning applications in the era of quantum technologies.
\end{abstract}

\maketitle
\end{CJK*}

\section{Introduction}
Quantum computation that leverages the principles of quantum mechanics has the potential to tackle problems that are beyond the reach of classical computers, 
revolutionizing fields ranging from cryptography~\cite{Pirandola2020} to finance~\cite{Herman2023} and drug discovery~\cite{Cao2018DrugDiscovery}.
Distributed quantum computing has attracted a lot of attention in recent years~\cite{Cuomo2020DQC,Caleffi2022distributed,Beals2013,Cacciapuoti2020,PhysRevLett.130.150602,montanaro2023quantum,gilboa2023exponential} due to the rapid progress in quantum communication technologies. In distributed quantum computing, multiple quantum processors are connected over a network, enabling collaborative computation and resource sharing.
This approach is crucial for scaling up quantum computing power and overcoming the limitations of individual quantum systems. Exploiting distributed quantum resources enables tackling larger and more computationally complex problems in domains such as optimization, simulation and quantum machine learning (QML). QML is especially suitable for distributed computation due to the need to process large datasets.

Privacy in distributed computing plays a vital role in ensuring the confidentiality and security of sensitive information processed by multiple parties.
Distributed quantum computation involves 
sharing and transmitting of quantum states across multiple nodes, 
making it paramount to protect the privacy of data and prevent unauthorized access. 
Furthermore, in practice, addressing privacy concerns in distributed quantum computing is essential for facilitating applications in fields such as finance and healthcare, where preserving the privacy of sensitive data is of utmost importance.


A number of protocols have been proposed in recent years that aim to implement private distributed quantum computing.
For example, blind quantum computing~\cite{Childs2005,Fitzsimons2017npjQI,PhysRevLett.111.230501} enables the client to execute a quantum computation using one or more remote quantum servers while keeping the structure of the computation hidden. 
Meanwhile, reducing the overhead in communication over blind quantum computation protocols has been an active research area since the first proposal of universal blind quantum computation (UBQC)~\cite{Childs2005}.
However, for distributed quantum computing problems such as QML, ensuring the privacy of data from a certain party while reducing the overhead in both quantum communication and computation remains a challenge.


In this work,  we 
introduce novel protocols for blind distributed quantum machine learning based on quantum bipartite correlator algorithm that can perform inner product estimation tasks.
Our protocols are communication-efficient compared with state-of-the-art classical and quantum blind distributed machine learning algorithms. Particularly, for the task of distributed inner product estimation,  a core subroutine in machine
learning applications, the protocols involve a communication complexity $O(\log N/\epsilon)$ with $N$ and $\epsilon$ being the size of the vectors and standard estimation error, respectively. We demonstrate how our protocols allow the client to conceal its data from the server, and vice versa.
We provide a detailed resource analysis for both communication and computation costs of our methods.
Our work paves the way for performing quantum machine learning with an untrusted device, while maintaining the privacy and keeping the resource overhead low.

\section{Formalism}

We start by presenting the problem statement in distributed quantum computation.
The basic setting includes two parties, Alice and Bob.
We assume that Alice has more quantum computational resources than Bob, such as a larger number of qubits. 
In many distributed quantum computation applications such as a delegated computation setting, Alice can be considered as a quantum server with Bob being a client. Furthermore, there is a quantum channel where qubits can be transmitted between the two parties. For the distributed QML tasks studied in this work, we assume that Alice holds the data $\boldsymbol{X}$ and Bob holds $\boldsymbol{y}$. For example, in supervised learning, $\boldsymbol{X}$ and $\boldsymbol{y}$ could be feature data and labels, respectively~\cite{Verbraeken2020}, while in unsupervised learning, both $\boldsymbol{X}$ and
$\boldsymbol{y}$ can be feature data with the objective to cluster them
based on distance estimation~\cite{barlow1989unsupervised}.

We consider the task of blind quantum machine learning, 
such as linear regression or classification~\cite{lloyd2013quantum,PhysRevLett.113.130503,li2019sublinear,Zhou2021BQML_clustering}. 
In machine learning, 
evaluating the inner product between two vectors is an important algorithmic building block.
The server holds the data vector $\boldsymbol{X}$ of size $N$  and the number of features for each data point is $M$, 
and the client holds a one-dimensional bitstring $\boldsymbol{y}$ with the same size $N$.
Note that transmitting the data classically to the server would introduce $O(N)$ complexity in communication. 
Meanwhile, as we consider distributed quantum computation, the data $\boldsymbol{X}$ and $\boldsymbol{y}$ are only held locally by the server and client, respectively.

In classical settings, the goal of achieving distributed machine learning with privacy can be approached using various techniques, such as homomorphic encryption~\cite{Fang2021HomorphicEncry,Wood2020HomorphicEncry}, which allows computation over encrypted data. Specifically, for distributed bipartite correlation estimation, many methods could be employed, including linearly homomorphic encryption~\cite{Paillier1999,Cheon2017}, non-interactive inner product protocols~\cite{Couteau2023Non-interactive} and oblivious-transfer-based secure computation~\cite{Boyle2020OT}.
However, it is important to note that these classical methods often introduce considerable overhead in terms of computation and communication complexity. Particularly, a communication cost of $\Tilde{O}(N)$ would be a minimum requisite~\cite{Couteau2023Non-interactive}.   As a result, their practical applications become limited, especially when dealing with large data sizes.

\section{Quantum bipartite correlator algorithm and its privacy}
In this section, we briefly introduce the quantum bipartite correlator (QBC) algorithm that can estimate the correlation between two bitstrings held by remote parties~\cite{PhysRevLett.130.150602}. The algorithm can be easily generalized to perform other computation tasks, such as the Hamming distance estimation. We remark that estimating bipartite correlation or Hamming distance serves as the building block of a general class of machine learning problems, including least-square fitting and classification of discrete labels~\cite{York1966,Tsoumakas2007}. 

Without loss of generality, we consider binary floating point numbers. We take the feature dimension
 $M$ 
to be one for simplicity hereafter unless specified. For two vectors $\boldsymbol{X}, \boldsymbol{y} \equiv [x_1,\cdots x_N]^T, [y_1,\cdots y_N]^T \in \{0,1\}^N$, 
we are interested in evaluating $ \overline{x y}=\frac{1}{N}\sum_{i=1}^{N}x_i y_i$ within a standard deviation error $\epsilon$. To begin with, we assume that the two parties Alice and Bob hold a local oracle that can encode their own data using 
a unitary transformation. That is, for Alice, one has $\hat{U}_{\vec{x}}: |i\rangle_n |0\rangle \mapsto |i\rangle_n |x_i\rangle$ that encodes the data $x_i$, where $|i\rangle_n$ is an $n\equiv\lceil\log_2(N)\rceil$-qubit (called index qubit hereafter) state $|i_1i_2\cdots i_n\rangle$, representing the index of the queried component with $i_k \in \{0,1\}$, $k \in [N]$, and $|x_i\rangle$ is a single-qubit state. Similarly, Bob has an oracle $\hat{U}_{\Vec{y}}$ of the same type that encodes his local data $y_i$. These oracle operators, as well as the ones introduced later, could be implemented with various techniques such as quantum random access memory~\cite{QRAM2008_PhysRevLett.100.160501}.

\begin{figure}[t]
\includegraphics[width=0.48\textwidth]{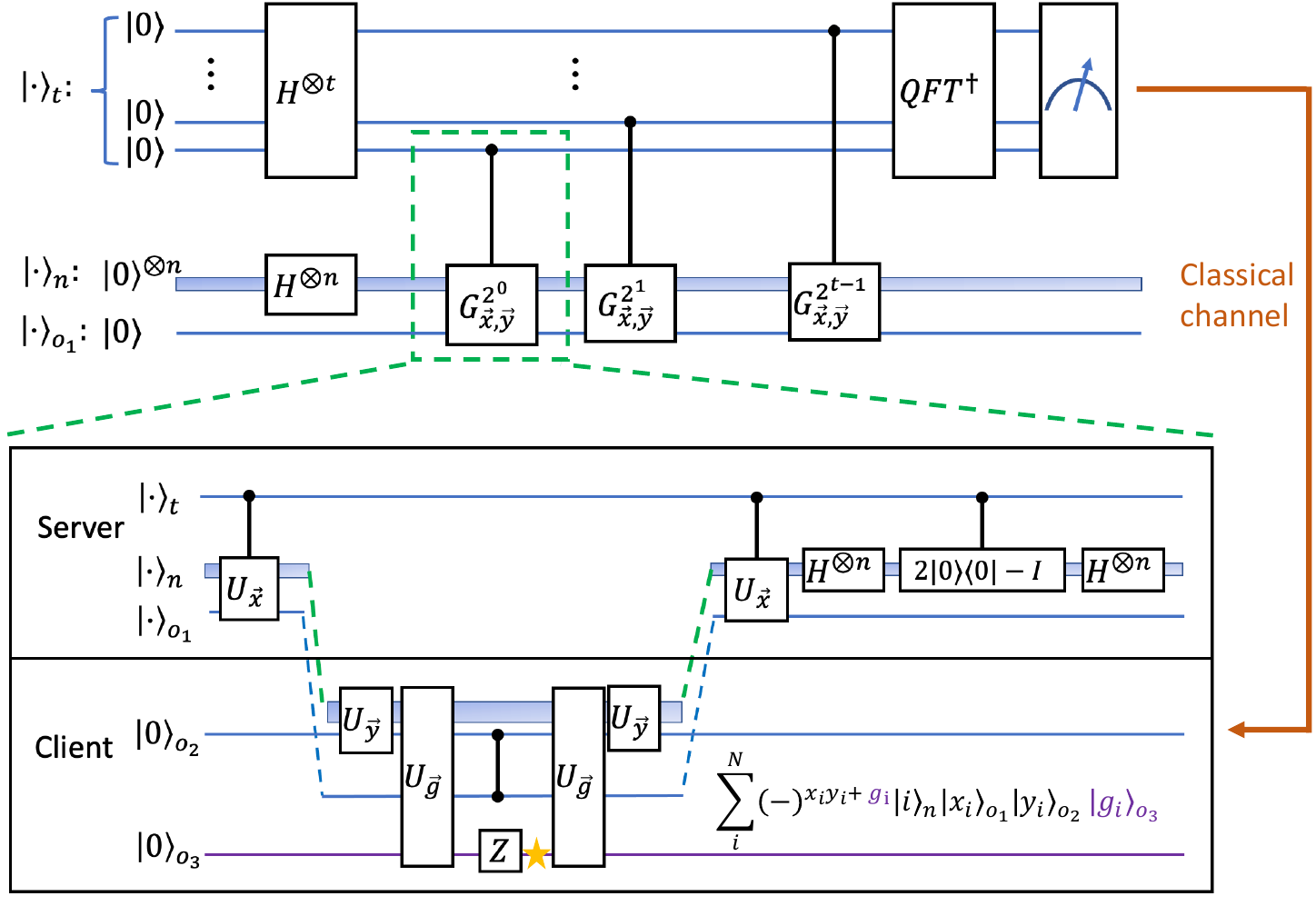}
\caption{Diagram for blind QBC with untrusted server. The upper diagram shows the quantum counting algorithm consisting Grover phase oracles $\hat{G}_{\vec{x},\vec{y}}$ and inverse QFT, while the lower box panel shows the realization details of each phase oracle. Compared to the original QBC algorithm, we introduce an ancillary qubit $o_3$ on client's side to add a phase $g_i$ during the computation process. The phase can be introduced via applying a phase gate on qubit $o_3$, which encodes a bitstring that is random and unknown to the server.
The detailed phase encoding rule is explained in the text. The quantum state at the star point is shown in the inset of the figure. After the server finishes the quantum circuit, it sends the extracted modified bipartite correlation $\frac{1}{N}\sum_i^N (x_i y_i +g_i)$ to the client via a classical communication channel. We omit the $1/\sqrt{N}$ normalization factor for index qubit states $\sum_i^N \ket{i}$ in the figures hereafter for simplicity.\label{fig: blindQBC_scheme} } 
\end{figure}

QBC is based on the quantum counting algorithm, where Alice and Bob send qubits via quantum channels and communicate with each other to realize the phase oracle~\cite{Brassard1998QuantumCounting,PhysRevLett.130.150602}, as shown in the top of Fig.~\ref{fig: blindQBC_scheme}. 
The quantum counting algorithm consists of a Grover operator $\hat{G}_{\vec{x},\vec{y}} \equiv \hat{H}^{\otimes n}(2|0\rangle_n \langle 0|_n-\hat{I})\hat{H}^{\otimes n} \hat{U}_{xy}$, where $\hat{U}_{xy}$ is a unitary operator that encodes information of both parties as we will introduce below, and inverse Quantum Fourier transform (QFT$^{\dagger}$) on register qubits $\ket{\cdot}_t$. When measuring the $t$-register, one can project it into a state $|j\rangle_t$ with phase $2\pi j\cdot2^{-t}$ which encodes either $\hat{\theta}$ or $2\pi-\hat{\theta}$, where $\theta = 2\arcsin{\sqrt{\overline{x y}}}$, with equivalent standard deviation: $\Delta \hat{\theta} = 2^{-t+1}$~\cite{PhysRevLett.130.150602}. 

During the phase oracle $\hat{G}_{\vec{x},\vec{y}}$, the following unitary circuit is applied to achieve encoding of $x_i$ and $y_i$
\begin{equation}\label{eq:phase_oracle_original}
  \hat{U}_{xy} \ket{i}_n \ket{00}_{o_1 o_2} = (-1)^{x_i y_i} \ket{i}_n\ket{00}_{o_1 o_2},
\end{equation}
where $o_1$, $o_2$ are two qubits locally held by Alice and Bob, respectively.
The above unitary operator can be implemented with the local oracles that Alice and Bob hold, i.e., $\hat{U}_{\Vec{x}}$ and $\hat{U}_{\Vec{y}}$. 

Specifically, Alice encodes her local information $\boldsymbol{X}$ into qubit $o_1$ via $\hat{U}_{\Vec{x}}$ operator and sends the $(n+1)$-qubit state $\frac{1}{\sqrt{N}}\sum_i^N \ket{i}_n \ket{x_i}_{o_1}$ to Bob via a quantum channel. After Bob applies his oracle and generates the state $\frac{1}{\sqrt{N}}\sum_i^N \ket{i}_n \ket{x_i}_{o_1} \ket{y_i}_{o_2}$, a controlled-Z (CZ) gate between qubit $o_1$ and $o_2$ is applied to encode the correlation information into the phase of the quantum state. That is, the bipartite quantum state is described by $\frac{1}{\sqrt{N}}\sum_i^N (-1)^{x_i y_i}\ket{i}_n \ket{x_i}_{o_1}\ket{y_i}_{o_2}$. The following local oracles would then yield the desired state $\frac{1}{\sqrt{N}}\sum_i^N (-1)^{x_i y_i}\ket{i}_n$ on which Alice will apply the quantum counting algorithm to estimate $ \overline{x y}=\frac{1}{N}\sum_{i=1}^{N}x_i y_i$ with bounded error $\epsilon$. We note that the CZ gate might be replaced with a different set of gates to estimate other types of correlations between $\boldsymbol{X}$ and $\boldsymbol{y}$. For example, to calculate their Hamming distance, one can implement the XOR gate $x_i \oplus y_i$ by replacing the CZ gate with a Z gate on $o_2$ sandwiched by two CNOT gates between $o_1$ and $o_2$~\cite{PhysRevLett.130.150602}. 


In the QBC algorithm, the communication complexity, i.e., the qubits transmitted during the overall process, is given by the Grover operation's $2(n+1)$ qubits communication repeated for $2^t-1$ iterations: 
\begin{equation}
	\mathcal C_{\rm comm} = 2(n+1)(2^t-1) = O\left(\frac{\log_2(N)}{\epsilon}\right),
	\label{eq:QBC_communication}
\end{equation}
where the number of register qubits $t$ is chosen to satisfy the desired error bound. 
We remark that the above communication complexity is advantageous compared with the 
SWAP-test-based algorithm that has a scaling of $O\left(\log_2(N)/\epsilon^2\right)$~\cite{PhysRevLett.124.060503} or LOCC-based algorithms with a scaling of $O\left(\log_2(N) \max \{ 1/ \epsilon^2, \sqrt{N} / \epsilon  \}\right)$~\cite{anshu2022distributed}.
This advantage is achieved by utilizing the distributed Grover operations.

The computational complexity, on the other hand, is the total number of oracle calls by Alice and Bob:
\begin{equation}
	\mathcal C_{\rm comp} = 4(2^t-1)=O\left(\frac{1}{\epsilon}\right).
	\label{eq:QBC_computation}
\end{equation}

We next consider the privacy of data in the QBC algorithm discussed above. From now on, we consider Alice as a server and Bob as a client. We first 
focus on the privacy of the client's information $\boldsymbol{y}$ to a semi-honest adversary. 
In this type of adversary, the honest-but-curious server follows the protocol and does not do any malicious behavior, but it tries to violate the privacy of the client's input by scrutinizing the messages transmitted in the protocol. That is, the server tries to infer $\boldsymbol{y}$ from the estimated $\frac{1}{N}\sum_{i}^{N}x_i y_i$.

In the trivial case when $x_i = 0, \forall i\leq N$, we have $\overline{x y}=0$ no matter what $\boldsymbol{y}$ is and the protocol has the best privacy. While in the worst case where 
the $x_i = 1, \forall i\leq N$ and $\overline{x y}=1$, the server could infer that $y_i = 1, \forall i\leq N$.  
In general, for $\boldsymbol{X}$ with Hamming weight $d_x$, the probability that the server gets the exact $\boldsymbol{y}$
 (that is, the Hamming distance between extracted and exact bitstring is $d_0=0$) 
is given by
\begin{equation}\label{eq:probablity_original_QBC}
  \text{Pr}(d_x) = \frac{1}{2^{N-d_x}} \frac{\prod_{i=1}^{d_x}i }{\prod_{i=1}^{N \overline{x y}}i \prod_{i=1}^{d_x - N \overline{x y} }i },
\end{equation}
where the factor $\frac{1}{2^{N-d_x}}$ comes from server having random guess on the indices $j$ that satisfies $x_j=0$. 
For a honest server in the original QBC protocol, however, the $\boldsymbol{y}$ information is always hidden from the server and is private.

In addition to the semi-honest adversary scenario discussed above, we note that in the original QBC algorithm, the preservation of privacy is not assured when we consider
a malicious server Alice. The server has the capability to acquire, to a certain extent, Bob's strings $\boldsymbol{y}$ by deviating from the expected quantum operations. We next discuss the designed blind QBC protocol with such an untrusted server.

\section{Blind QBC with untrusted server}\label{sec:protocol_untrusted_server}

A malicious server can get the client's information by deviating from the established QBC protocol. 
One example is that the server could perform quantum gate operations and measurements to extract the phase information instead of following the expected Grover steps after receiving $\frac{1}{\sqrt{N}}\sum_i^N (-1)^{x_i y_i}|i\rangle_n|x_i\rangle_{o_1}$ from the client Bob. Alternatively, a malicious server could potentially manipulate the state of qubit $o_1$ sent to the client, rather than genuinely encoding the information of $\boldsymbol{X}$. In principle, for each communication round, the server can acquire one bit of information of client's data $\boldsymbol{y}$.
Then with the $2^t-1 = O(\frac{1}{\epsilon})$ Grover iterations, the server could get $O(\frac{1}{\epsilon})$ bits of information in $\boldsymbol{y}$.
Such an attack strategy might be implemented by preparing the $o_1$ qubit in $\ket{+}$ state and sending $\frac{1}{\sqrt{N}}\sum_i^N \ket{i}_n\ket{+}_{o_1}$ to the client (Appendix~\ref{app:attack_strategy}). Subsequent to the reception of the quantum state from the client, the server undertakes an $X$ basis measurement on qubit $o_1$. The server could  perform the sampling procedure encompassing the bitstrings of the index qubits during the $O(\frac{1}{\epsilon})$ communication rounds. 

We note that the server could not manipulate the index qubit states  $\frac{1}{\sqrt{N}}\sum_i^N \ket{i}$ to amplify the amplitude of a specific bistring of interest, as the client is capable of verifying the received quantum state of index qubits by performing X basis measurements to check whether they have the same amplitude. On the other hand, it is possible to employ a redundant encoding strategy to further decrease the probability that the server attains a specific $y_i$ corresponding to an intended index. However, this comes at the expense of increased communication complexity, as detailed in Appendix.~\ref{app: redundant_encoding}.

To counteract the aforementioned attack strategy, we need to devise a protocol enabling the server to execute machine learning tasks  while remaining unaware of the exact label information $\boldsymbol{y}$, even when the malicious server does not follow the designed protocol. 
In this case, we consider an honest client, who is not interested in learning $\boldsymbol{X}$. This assumption might be removed if we consider further encoding privacy in $\boldsymbol{X}$ when sending information to the client. 
To implement remote blind bipartite correlation estimation, a desired protocol should have 1) less overhead in quantum communication, 2) less requirements in the computational power of client, 3) a certified estimation result with error $\epsilon$. 

We thus consider the revised QBC algorithm below (Fig.~\ref{fig: blindQBC_scheme}). Inspired by quantum one-time pad~\cite{Childs2005}, the protocol utilizes phase padding to preserve privacy.
The client Bob now has one or more qubits at hand, where he can encode a bit string $\ket{g_i}$ that is blind to the server. That is, the client has an oracle $\hat{U}_{\Vec{g}}$ for the extra qubit (denoted as $o_3$ hereafter), and the modified phase oracle of Eq.~\ref{eq:phase_oracle_original} reads as
\begin{equation}
   \hat{U}_{xyg} \ket{i}_n \ket{000}_{o_1 o_2 o_3} = (-1)^{x_i y_i+g_i} \ket{i}_n\ket{000}_{o_1 o_2 o_3}.
\end{equation}
To implement the above unitary $\hat{U}_{xyg}$, similar to the $\hat{U}_{xy}$, the client performs $\hat{U}_{\Vec{y}}$ and $\hat{U}_{\Vec{g}}$ oracle after receiving state from server to create the state $\frac{1}{\sqrt{N}}\sum_i^N \ket{i}_n \ket{x_i}_{o_1} \ket{y_i}_{o_2} \ket{g_i}_{o_3}$, followed by a controlled-Z gate between $o_1$ and $o_2$. Then a local Z gate can be applied on qubit $o_3$ to add the phase $(-1)^{g_i}$ that is random to server.

Since the phase term  $(-1)^{x_iy_i+g_i}$ is binary here with modular addition between $x_i y_i$ and $g_i$, we design the following rule for the application of random phase $g_i$. For a given index $i$, when $y_i=0$, the client chooses a random number from $\{0,1\}$; 
while when $y_i=1$, the client sets $g_i=0$. 
Under this setting, the server cannot get $y_i$ in general from direct measurement of the parity at each Grover step, even if the server knows exactly the circuit that the client performs. 

The above phase encoding rule on $g_i$ guarantees that $x_i y_i + g_i \in \{0,1 \}$. The quantum counting algorithm can then estimate $\frac{1}{N}\sum_i^N (x_i y_i + g_i) = \frac{1}{N}\sum_i^N (x_i y_i + g_i \mod 2)$ 
with error bound $\epsilon$. Finally, after the measurement, the server sends the estimated result back to the client via a classical channel, from which the client can extract $\frac{1}{N}\sum_i^N x_i y_i$ using his local information of $\frac{1}{N}\sum_i^N g_i$. Alternatively, depending on the specific use cases, the client could directly share $\frac{1}{N}\sum_i^N g_i$ with the server and let it extract the bipartite correlation between $\boldsymbol{X}$ and $\boldsymbol{y}$.

We emphasize that in principle, the aforementioned protocol could still inadvertently leak a portion of the information in $\boldsymbol{y}$ to the server. As can be seen from the scheme, in the case where $x_j=1$ and the final phase term is $x_j y_j + g_j = 0$, if the server knows the above application rule of $g_i$ and extracts the phase corresponding to the index qubit $\ket{i}_{i=j}$, it could infer that $y_j=0$. We consider the worst scenario where the malicious server picks $x_i = 1, \forall i \leq N$ and has client's local phase encoding rule. The server's attack strategy is to measure the phase of a randomly picked index $\ket{i}$ to extract $x_i y_i + g_i$ at each Grover iteration. Then, for $\boldsymbol{y}$ with Hamming weight $d_y$, the probability that the server extracts a bitstring $ \boldsymbol{y^{\prime}}$ that is $d_0$-close ($d_0\leq d_y$) to $\boldsymbol{y}$ using the information of the measured phases and without doing random guess is simply given by
\begin{equation}\label{eq:probability_untrusted_server}
\begin{split}
  & \text{Pr}(d(\boldsymbol{y}, \boldsymbol{y^{\prime}}) = d_0) = \\
  & \frac{ C(d_y, d_0 ) C(N-d_y, \min(2^t-1, d_y)-d_0)}{C(N, \min(2^t-1, d_y))} 
\end{split}
\end{equation}
where $C(\cdot,\cdot)$ denotes the binomial coefficient. 
As can be seen from the analysis above, even in the worst case, the probability that the server can successfully extract part of $\boldsymbol{y}$ information becomes considerably low when the data size becomes large, particularly when $N \geq 2^t-1$, while in the original QBC a malicious server could get $2^t-1$ bits of information from the client during the communication round.
Note that the iteration number $2^t-1$ yields the standard deviation of the estimated correlation, that is, $2^t-1 = O(\frac{1}{\epsilon})$. A less tight error bound $\epsilon$ will reduce the number of communication rounds between server and client thus increasing the privacy of client's data. 

We remark that the quantum communication complexity of the aforementioned algorithm for blind server is $\mathcal C_{\rm comm}^{b_s} = O(\frac{\log_2(N)}{\epsilon})$, which is the same as the original QBC as depicted in Eq.~\ref{eq:QBC_communication}. 
Moreover, akin to the QBC algorithm, a classical communication channel is needed at the end of QBC to deliver estimation results to the client. In terms of computational overhead experienced by the client, introducing the ancilla qubit $o_3$ only adds $O(\frac{1}{\epsilon})$ number of two-qubit phase gates and as a result, does not alter the inherent computational complexity. To this end, the blind QBC protocol proposed here could enable communication-efficient blind distributed machine learning tasks between a server and a client without presupposing substantial quantum resources on the client.

\section{Blind QBC with untrusted client}\label{sec: untrusted_client}

\begin{figure*}[t]
\centering
\includegraphics[width=0.86\textwidth]{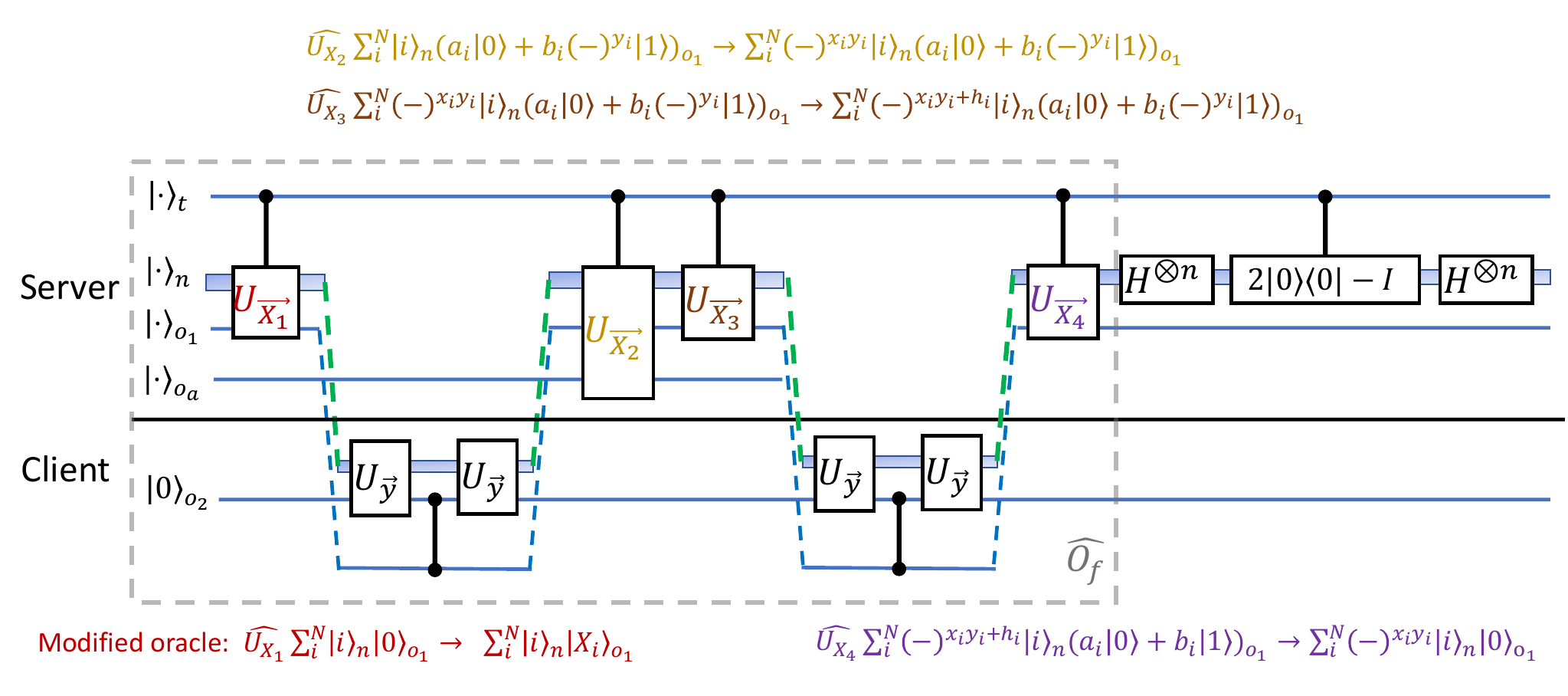}
\caption{Grover operator $\hat{H}^{\otimes n}(2|0\rangle_n \langle 0|_n-\hat{I})\hat{H}^{\otimes n} \hat{O}_{f}$ for  blind quantum bipartite correlator protocol to hide server data $\boldsymbol{X}$ from client. The operator starts with an oracle held by server (Alice) that encodes $\boldsymbol{X}$ with random basis (oracle $\hat{U}_{X_1}$). After receiving the state returned by client (Bob), the server extracts the desired phase term $(-1)^{x_i y_i}$ ($\hat{U}_{X_2}$) and return an encoded state back to client  ($\hat{U}_{X_3}$) to remove the phase in $o_1$ qubit that the server does not know. Finally, the server reaches the target state $\frac{1}{\sqrt{N}}\sum_i^N (-1)^{x_i y_i} \ket{i}_n$ by decoupling $o_1$ qubit with index qubits  ($\hat{U}_{X_4}$). \label{fig: blindClient_scheme} }
\end{figure*}

We now discuss the scenario where the server would like to estimate $\frac{1}{N}\sum_i^N x_i y_i$ while keeping $\boldsymbol{X}$ hidden from the client at all times during the process. In practical applications such as model-as-a-service platforms~\cite{MaaS2018,Hesamifard2018}, the server's information, including the model's parameters or training data, should remain hidden from the clients. 
By hiding the server-side information, they can prevent the client from reverse-engineering or extracting valuable information about the underlying model architecture or training data. 
Under this setting, the protocol should be secure against not only a honest-but-curious client, but also a malicious client who tries to get $\boldsymbol{X}$ by deviating from the original quantum algorithm. 

Here we assume an honest server that follows the protocol exactly without trying to get the label information $\boldsymbol{y}$. The goal is then to encode $\boldsymbol{X}$ when the server sends qubits to the client while running the QBC algorithm. That is, we are interested in designing a privacy-preserving operator $\hat{O}_f$ such that
\begin{equation}
  \hat{O}_f \frac{1}{\sqrt{N}}\sum_i^N \ket{i}_n \ket{00}_{o_1 o_2} = \frac{1}{\sqrt{N}}\sum_i^N (-1)^{x_i y_i}\ket{i}_n \ket{00}_{o_1 o_2}.
\end{equation}
Inspired by quantum key distribution protocols~\cite{QKD_RevModPhys.92.025002} such as BB84~\cite{Bennett2014},
we consider a modified local oracle operator $\hat{U}_{X_1}$ held by the server, where the data information $\boldsymbol{X}$ is encoded in different basis (Fig.~\ref{fig: blindClient_scheme}). Specifically, at each iteration of quantum counting algorithm, for a given index $i$, the server chooses a random number $R_i$ from $\{0,1 \}$. When $R_i=0$, the server encodes $x_i$ using the Z basis, i.e., $\ket{i}_n\ket{0}_{o_1}$ or $\ket{i}_n\ket{1}_{o_1}$, depending on whether $x_i$ being $0$ or $1$; if $R_i=1$, $x_i$ is encoded in the X basis and now the state reads $\ket{i}_n\ket{+}_{o_1}$ or $\ket{i}_n\ket{-}_{o_1}$. Here $\ket{+(-)}=\frac{1}{2} (\ket{0}\pm \ket{1})$ are the eigenstates of Pauli X operator. This oracle $\hat{U}_{X_1}$ can be implemented with the original oracle $\hat{U}_{\Vec{x}}$ with Hadamard gates on $o_1$ conditioned on index $\ket{i}_n$. 

Then, the state received by the client at each time reads as $\frac{1}{\sqrt{N}}\sum_i^N \ket{i}_n\ket{X_i}_{o_1}$ with $X_i$ being $1(0)$ or $+(-)$. 
As the client does not know which basis the server chooses for given $i$, at each Grover iteration, measurement of qubit $o_1$ on index $\ket{i}$ will have the probability of yielding both 0 or 1, hence the client cannot infer the $x_i$ information from the single copy of the received $\frac{1}{\sqrt{N}}\sum_i^N \ket{i}_n\ket{X_i}_{o_1}$ state. Note that the server could pick different random numbers $R_i$ at different communication rounds when executing the QBC algorithm.

 As in the original QBC algorithm, the client performs CZ gate between the received qubit $o_1$ and local qubit $o_2$ sandwiched by $\hat{U}_{\vec{y}}$ operators. Then, the state received by the server from the quantum channel is $\frac{1}{\sqrt{N}}\sum_{i}^N \ket{i}_n (a_i\ket{0}+b_i (-1)^{y_i}\ket{1})_{o_1}$ where $a_i(b_i)$ is decided by $x_i$ and the encoding basis $R_i$ thus is known to the server. 
 We next discuss how the server could perform operations to reach  the target state $\frac{1}{\sqrt{N}}\sum_i^N (-1)^{x_i y_i}\ket{i}$ for running the follow-up QBC algorithm.
 We consider a second oracle operator held by the server $\hat{U}_{X_2}$:
 \begin{equation}\label{eq:U_X_2}
 \begin{split}
    & \hat{U}_{X_2}\frac{1}{\sqrt{N}}\sum_{i}^N \ket{i}_n (a_i\ket{0}+b_i (-1)^{y_i}\ket{1})_{o_1} = \\
    & \frac{1}{\sqrt{N}} \sum_{i}^N (-1)^{x_i y_i}\ket{i}_n (a_i\ket{0}+b_i (-1)^{y_i}\ket{1})_{o_1}.
\end{split}
 \end{equation}
This can be achieved via the help of an additional qubit $o_a$ held by the server that encodes the $\boldsymbol{X}$ information in the normal Z basis (see Appendix~\ref{app: U_X_2} for details of circuit implementation). 

Note that the server cannot decouple the $o_1$ qubit with an unknown state, as the honest server only has the information of $a_i$ and $b_i$ but doesn't have the information of $\boldsymbol{y}$. In order to reset the state of qubit $o_1$, the server could return the state back to client to have the client remove the phase $(-1)^{y_i}$. Before doing so, the server would like to first hide its information by adding a random phase padding by applying $\hat{U}_{X_3}$ which is defined as
 \begin{equation}\label{eq:U_X_3}
 \begin{split}
    & \hat{U}_{X_3}\frac{1}{\sqrt{N}} \sum_{i}^N (-1)^{x_i y_i}\ket{i}_n (a_i\ket{0}+b_i (-1)^{y_i}\ket{1})_{o_1} = \\
    &\frac{1}{\sqrt{N}} \sum_{i}^N (-1)^{x_i y_i + h_i}\ket{i}_n (a_i\ket{0}+b_i (-1)^{y_i}\ket{1})_{o_1}.
\end{split}
 \end{equation}
Here $h_i \in \{0,1 \}$ is blind to the client and could change in different communication rounds, therefore the client would not be able to extract $x_i$ information. 
The client performs a controlled-Z gate again between its local qubit $o_2$ and the received qubit $o_1$, after which the phase term $(-1)^{y_i}$ becomes $(-1)^{y_i+y_i} = 1$. Then, the server receives the state $\frac{1}{\sqrt{N}} \sum_{i}^N (-1)^{x_i y_i + h_i}\ket{i}_n (a_i\ket{0}+b_i \ket{1})_{o_1}$ from client and performs oracle $\hat{U}_{X_4}$:
 \begin{equation}\label{eq:U_X_4}
 \begin{split}
    & \hat{U}_{X_4}\frac{1}{\sqrt{N}} \sum_{i}^N (-1)^{x_i y_i + h_i}\ket{i}_n (a_i\ket{0}+b_i \ket{1})_{o_1} = \\
    &\frac{1}{\sqrt{N}} \sum_{i}^N (-1)^{x_i y_i}\ket{i}_n \ket{0}_{o_1} .
\end{split}
 \end{equation}
It can be easily seen that to implement $\hat{U}_{X_4}$, the server could simply perform $\hat{U}_{X_3}$ again to remove the added random phase term $(-1)^{h_i}$ and then reset the qubit $o_1$ to $\ket{0}_{o_1}$ as the server knows the all coefficients $a_i$ and $b_i$. 

We remark that the random numbers $R_i$ and $h_i$ can change in different Grover iterations. That is, the client will not get useful information by performing measurements on each iteration and using the joint results from a sequence of measurements to infer $\boldsymbol{X}$. The privacy of $\boldsymbol{X}$ is guaranteed by the fact that measuring a single copy in a given basis cannot reveal both the basis information $R_i$ and the data information $x_i$. The probability that the client gets $\boldsymbol{X^{\prime}}$ that is $d_0$-close to the true $\boldsymbol{X}$ would simply be the same as a random guess.

To this end, we have described a phase encoding oracle $\hat{O}_f$ that lets the server acquire the state $\frac{1}{\sqrt{N}}\sum_i^N (-1)^{x_i y_i}\ket{i}_n$ for subsequent operations without leaking the information of data $\boldsymbol{X}$ to an untrusted client. The scheme is based on a random encoding of $\boldsymbol{X}$ and is information-theoretic secure against an untrusted client, with the proof of security following directly from the corresponding proof for the BB84 protocol~\cite{Bennett2014,PhysRevA.72.012332}. The total number of oracle calls by server and client only increases by a constant at each iteration, thus leading to the same computation complexity $O(\frac{1}{\epsilon})$ as Eq.~\ref{eq:QBC_computation}. The total communication cost of this blind client scheme is given by 
\begin{equation}
	\mathcal C_{\rm comm}^{b_c} = 4(n+1)(2^t-1) = O\left(\frac{\log_2(N)}{\epsilon}\right),
	\label{eq:untrusted_client_communication}
\end{equation}
which has the same complexity scaling as the original QBC algorithm. We summarize the proposed algorithms here and above in Table.~\ref{table:alg_summary}.

\begin{table*}
\centering
\caption{Privacy and communication complexity of proposed distributed inner product estimation algorithms.\label{table:alg_summary}}
\begin{tabular}{  m{8em} | m{4cm}| m{3.5cm} |m{4cm} | m{2.8cm} } 
  \hline \hline
  Adversaries & Protocol & Privacy mechanism &Privacy & Communication complexity \\ 
  \hline
  Honest-but-curious server & original QBC algorithm~\cite{PhysRevLett.130.150602} & - & worst scenario in Eq.~\ref{eq:probablity_original_QBC} & $O((\log_2 N)/\epsilon)$ \\ 
  \hline
  Malicious server & blind QBC for untrusted server (Sec.~\ref{sec:protocol_untrusted_server})& random phase padding & worst scenario in Eq.~\ref{eq:probability_untrusted_server}& $O((\log_2 N)/\epsilon)$\\ 
  \hline
  Malicious client & blind QBC for untrusted client (Sec.~\ref{sec: untrusted_client}) & random basis encoding, random phase padding & information-theoretic secure & $O((\log_2 N)/\epsilon)$\\ 
  \hline
  \hline
\end{tabular}
\end{table*}

\section{Generalization into multi-party settings}
The algorithms discussed above can be generalized into multi-party settings and find applications in secure multi-party computation and machine learning~\cite{crepeau2002secure,MPC_ML2021}, where parties collaboratively perform computations on their combined data sets without revealing the data they possess to untrusted parties. For example, to perform model aggregation, an untrusted central server would like to perform linear regression or classification using its local data as well as labels that are distributed among multiple clients. Then, the protocol in Sec.~\ref{sec:protocol_untrusted_server} can be applied in which the server can interact with each client to extract model parameters individually. 

Here we provide an example of multi-party protocols. We consider a system consisting of a central server and $m$ clients, where the server is untrusted by the clients. The task  is to have the server evaluate $f_m = \frac{1}{N}\sum_i^N (\sum_j^m x_i y_i^{(j)} \mod 2)$ without leaking individual information of clients. Similar to the phase pad technique introduced in Sec.~\ref{sec:protocol_untrusted_server}, one can protect each individual client's information by adding additional terms in the phase when running the QBC algorithm. Specifically, we consider a cascaded protocol where each client encodes its information into the phase of index qubits and passes the state into the next client. In each communication round, the $k$-th client would receive the state $\frac{1}{\sqrt{N}}\sum_i^N (-1)^{x_i y_i^{(1)} + x_i y_i^{(2)} + ... + x_i y_i^{(k-1)}} \ket{i}_n\ket{x_i}$ from the $(k-1)$-th client. Then, by applying CZ gate between $o_1$ and its local qubit, the $j$-th client  sends the state $\frac{1}{\sqrt{N}}\sum_i^N (-1)^{x_i y_i^{(1)} + x_i y_i^{(2)} + ... + x_i y_i^{(k-1)} +  x_i y_i^{(k)}} \ket{i}_n\ket{x_i}$ to the next client.  The final $m$-th client will pass the state $\frac{1}{\sqrt{N}}\sum_i^N (-1)^{\sum_j^m x_i y_i^{(j)}} \ket{i}_n\ket{x_i}$ to the server which can then perform the remaining part of the original QBC algorithm to extract the desired $f_m$. 

We note that a malicious server could only get the $\sum_j^m y_i^{(m)}$ and the individual $y_i^{(j)}$ information is not leaked, as the phase added by each client servers as a random pad of other clients. For the same reason, the $j$-th ($j\geq 3$) client cannot get previous clients' information as it can only extract $\sum_{k=1}^{j-1} y_i^{(k)}$. The first client ($j=1$) can further add a random pad $g_i^{(1)}$ to protect its information against the second client ($j=2$). 
The protocol here is similar to incremental learning~\cite{Sheller2020}, where the model aggregation is performed while preserving privacy. We remark that the total communication cost scales as $O\left(\frac{m\log_2(N)}{\epsilon}\right)$ and the privacy mechanism does not introduce additional communication cost. 
To this end, our work paves the way for communication-efficient private machine learning for multi-party system, such as quantum federated learning~\cite{QFL_Li2021,QFL_Larasati2022,kumar2023expressive}.

\section{Discussion and conclusion}

As mentioned above, the proposed blind distributed inner product estimation protocols can be applied in distributed machine learning where a central task is to evaluate correlations between remote matrices or vectors. Here we give an example of such applications.  In linear regression problems, one is interested in finding the coefficient vector $\boldsymbol{\lambda}$ with standard error $\epsilon$ that satisfies $\boldsymbol{X}_{N\times M}\boldsymbol{\lambda}_{M\times 1} = \boldsymbol{y}_{N\times 1}$, where the $N$-by-$M$ matrix $\boldsymbol{X}$ and $N$-by-1 vector $\boldsymbol{y}$ are separately held by two remote parties, a server and a client, respectively. We consider the case where the server would like to estimate  $\boldsymbol{\lambda}$ without letting the client extract its local information $\boldsymbol{X}_{N\times M}$.
The $l$-th component of $\boldsymbol{\lambda}$ reads $\lambda_l = \sum_{i=1}^N X_{li}^{\dagger}y_i$, where $l$ and $i$ labels the index of the element in the matrix or vector.
The problem can be reduced to estimate product of distributed numbers $a_{li}=X^{\dagger}_{li}$ and $b_i=y_i$. They can be expanded as binary floating point
numbers using, for example, $a_{li} = \sum_{k=0}^{\infty} 2^{u-k}x^{(k)}_{li}$ and $b_i = \sum_{k=0}^{\infty} 2^{v-k}y^{(k)}_{i}$, for which $u$ and $v$ denote the highest digits of $a$ and $b$, respectively~\cite{PhysRevLett.130.150602,IEEEstandard2019}. Then, the target coefficient $\lambda_l$ can be written as $\lambda_l = \sum_{i=1}^N a_{li} b_i = \sum_{r=0}^{\infty} 2^{u+v-r} \sum_{k=0}^r \sum_{i=1}^N x^{(k)}_{li} y^{(r-k)}_{i} $, where the blind QBC algorithm introduced in Sec.~\ref{sec: untrusted_client} can be directly applied. In this case, the untrusted client can neither directly extract the information of $\boldsymbol{X}_{N\times M}$ during the blind QBC communication, nor indirectly have an estimation on $\boldsymbol{X}_{N\times M}$ from the knowledge of coefficient $\boldsymbol{\lambda}_{M\times 1}$.
To this end, our proposed algorithms exhibit direct applicability within the domain of distributed blind machine learning tasks, particularly in scenarios involving matrix or vector multiplication operations.

We further remark that the proposed quantum algorithms offer many benefits for practical applications with large data sizes. Notably, the quantum communication cost in estimating the bipartite correlation scales as $O(\frac{\log N}{\epsilon})$ and additionally, the discussed data privacy mechanism does not impose any additional overhead in terms of communication cost. Furthermore, the protocols eliminate the need for a trusted third party and necessitate only a minimal quantum resource allocation from the participating clients, encompassing the number of qubits and gate operations.

In summary, this study introduces novel blind quantum machine learning protocols that utilize a quantum bipartite correlator estimation algorithm for distributed parties. By addressing the potential threat of malicious parties attempting to extract information from others, we propose two distinct settings that ensure privacy preservation for each party in the QBC algorithm. Leveraging the advantageous properties of quantum phases and the flexibility of encoding data in various bases, our protocols can effectively safeguard information. The developed blind QML algorithm offers notable advantages, including low communication and computational complexity. This work contributes to the advancement of secure and efficient QML protocols, thus presenting an efficient pathway for distributed quantum computing.

\acknowledgements
JL acknowledges support by DTRA (Award No. HDTRA1-20-2-0002) Interaction of Ionizing Radiation with Matter (IIRM) University Research Alliance (URA).

\section*{Disclaimer}
This paper was prepared for informational purposes with contributions from the Global Technology Applied Research center of JPMorgan Chase $\&$ Co. This paper is not a product of the Research Department of JPMorgan Chase $\&$ Co. or its affiliates. Neither JPMorgan Chase $\&$ Co. nor any of its affiliates makes any explicit or implied representation or warranty and none of them accept any liability in connection with this position paper, including, without limitation, with respect to the completeness, accuracy, or reliability of the information contained herein and the potential Legal, compliance, tax, or accounting effects thereof. This document is not intended as investment research or investment advice, or as a recommendation, offer, or solicitation for the purchase or sale of any security, financial instrument, financial product or service, or to be used in any way for evaluating the merits of participating in any transaction.

\newpage
\clearpage

\appendix
\section{Extraction of $y_i$ information in QBC by malicious server}\label{app:attack_strategy}
We discuss a feasible attack protocol for a malicious server to extract information of $\boldsymbol{y}$ with the received state $\sum_i^N (-1)^{x_i y_i} \ket{i}_n \ket{x_i}$ in the original QBC algorithm.
In this protocol, the server prepares the $o_1$ qubit simple in the $\ket{+}=\frac{1}{\sqrt{2}}(\ket{0}+\ket{1})$ state. The quantum state sent to client would then be
\begin{equation}
  \frac{1}{\sqrt{2}}( \frac{1}{\sqrt{N}}\sum_i^N \ket{i}_n \ket{0}_{o_1} + \frac{1}{\sqrt{N}} \sum_i^N \ket{i}_n \ket{1}_{o_1})
\end{equation}
The honest client then encodes $y_i$ information in the phase with his own local oracle, leading to state
\begin{equation}\label{eq:extraction_state}
\begin{split}
  &  \frac{1}{\sqrt{2N}}(  \sum_i^N \ket{i}_n \ket{0}_{o_1} + \ket{1}_{o_h} \sum_i^N (-1)^{y_i}\ket{i}_n \ket{1}_{o_1}) \\
  & = \frac{1}{\sqrt{2N}} \sum_i^N \ket{i}_n (\ket{0}_{o_1} + (-1)^{y_i} \ket{1}_{o_1} )
\end{split}
\end{equation}
that is sent back to server.

Then, it's clear to see that to extract client's information, the server could perform measurement on qubit $o_1$ in the X basis and extract the $y_j$ information depending on the measured index qubit bitstring $j$. In this case, by performing sampling on the N index qubit states during the $2^t-1 = O(1/\epsilon)$ communication rounds, the malicious server could get $O(1/\epsilon)$ information of $\boldsymbol{y}$. 
Indeed, given the state Eq.~\ref{eq:extraction_state} received by the server, the upper bound of information that the server could get at each round by performing measurement on index qubits and qubit $o_1$ is determined by the Holevo's bound~\cite{Holevo1973bounds}:
\begin{equation}
\begin{split}
    &H(C:S) \leq S(\rho) - \frac{1}{N} \sum_i^N S(\rho(i))  = \log 2N,
    \end{split}
\end{equation}
where $S(\rho)$ denotes the von Neumann entropy for density matrix $\rho$ that corresponds to Eq.~\ref{eq:extraction_state}, and $\rho_i = \ket{a_i}\ket{i}\bra{i}\bra{a_i}$ ($a_i = +, -$) forms the POVM set that server performs. 

One might argue that the server could amplify the probability of sampling a particular index qubit bitstring $j$ by reducing the amplitude of other index qubit bitstrings.  That is, the quantum state sent to client could be
\begin{equation}\label{eq:non-equal-superposition}
  \frac{1}{\sqrt{2}}( \sum_i^N A_i \ket{i}_n \ket{0}_{o_1} +  \sum_i^N A_i \ket{i}_n \ket{1}_{o_1})
\end{equation}
where $|A_{i=j}|^2 \gg |A_{i\neq j}|^2$ and $\sum_i^N |A_i|^2 =1$. 
However, the client can add an additional verification on the $\lceil\log_2(N)\rceil$ index qubits upon receiving them by performing measurements on X basis. This should yield $+1$ for all index qubits, as the state $\frac{1}{\sqrt{N}}\sum_i^N \ket{i}$ can be rewritten as $\frac{1}{\sqrt{2}}(\ket{0}+\ket{1})^{\otimes \lceil\log_2(N)\rceil }$. While for the manipulated state outlined in Eq.~\ref{eq:non-equal-superposition}, there exists a nonzero probability of producing a measurement outcome of $-1$ for at least a portion of the measurements.

\section{Redundant encoding against malicious server}\label{app: redundant_encoding}
We describe a redundant encoding approach aimed at reducing the probability that a malicious server acquiring a specific $y_i$ information with $i$ being the pertinent index of interest using the attack strategy in Appendix.~\ref{app:attack_strategy}.

Given that the server is restricted to preparing the index qubits in a manner where each index bitstring holds identical probability, after receiving the state back from client, the probability that server samples a specific index bitstring $\ket{i}_n$ is simply $\frac{1}{N}$. That is, in each iteration of communication during the execution of QBC algorithm, the server is constrained to attain a specific $y_i$ corresponding to the intended index with a probability of $\frac{1}{N}$; and for $1/\epsilon$ iterations needed for QBC algorithm, this will cause a total amount of information being extracted to be $\frac{1}{N\epsilon}$. Following this, we can consider a protocol where both the client and server encode their single bit local information ${y_i}$ and ${x_i}$ into bitstrings ${\left[y^{\prime}_{i,1},y^{\prime}_{i,2},\cdots,y^{\prime}_{i,M}\right]}^{\prime}$ and ${\left[x^{\prime}_{i,1},x^{\prime}_{i,2},\cdots,x^{\prime}_{i,M}\right]}$ with size $M$, where $M>1$. The total amount of bits increases from $N$ to $MN$. The encoding rule is shown as follows:
\begin{equation}\label{eq:rule_extending_bitstring_size_X}
    \begin{split}
        \boldsymbol{x}^{\prime}_{i,j} = x_i; \quad i = 1,2,...,N; j = 1,2, ..., M;
    \end{split}
\end{equation}
which is a simply copy the bit $x_i$ for $M$ times. As for $\boldsymbol{y}^{\prime}$, the client can hide the information $y_i$ randomly in one of the $M$ digits and let the other $M-1$ digits to be all zero or one. That is, client chooses either
\begin{equation}\label{eq:rule_extending_bitstring_size_y1}
    \begin{split}
        &\boldsymbol{y}^{\prime}_{i,j} = \delta_{j, J_i} \cdot y_i; \\
         i = 1,2,...,N; &j = 1,2, ..., M, J_i\in\{1,2,...,M\}. \\
    \end{split}
\end{equation}
or 
\begin{equation}\label{eq:rule_extending_bitstring_size_y2}
    \begin{split}
         &\boldsymbol{y}^{\prime}_{i,j} = (1-\delta_{j, J_i}) \cdot y_i; \\
         i = 1,2,...,N; &j = 1,2, ..., M, J_i\in\{1,2,...,M\}. \\
    \end{split}
\end{equation}
where $J_i$ is an random number and $\delta_{j, J_i}$ is the Kronecker symbol. 
In these cases, the server would get $\frac{1}{NM}\sum_i^{N} x_i y_i$ or $\frac{1}{NM}\sum_i^{N} x_i y_i + \frac{M-1}{NM}\sum_i^N x_i$ by executing the QBC algorithm,  depending on whether the client chooses encoding method Eq.~\ref{eq:rule_extending_bitstring_size_y1} or Eq.~\ref{eq:rule_extending_bitstring_size_y2}.
Afterwards, the client can send an one-bit message via classical channel to the server and let server knows which one was used.

We remark that at each communication round, the probability that the server samples a specific bit reduces from $\frac{1}{N}$ to $\frac{1}{NM}$. Even though that $M$-times more communication round will be needed to achieve the same error bound $\epsilon$ as in the original QBC case, the server would not know which digit encodes the correct $y_i$ information as here $J_i$s are random numbers. Therefore, using the attack strategy detailed in the Appendix.~\ref{app:attack_strategy}, the probability that the server successfully gets a specific bit $y_i$ would be $\frac{1}{NM} \times \frac{M}{\epsilon} \times \frac{1}{M} = \frac{1}{NM \epsilon}$, where the second term $\frac{M}{\epsilon}$ is the total number of communication rounds and the third term is $\frac{1}{M}$ is due to the randomness in $J_i$. 
 It's clear to see that a larger value of $M$ corresponds to a decreased probability for the server to successfully extract valuable information from the client through the attack strategy. The flexibility that the client can independently choose encoding method also protects the majority information of $\boldsymbol{y}$, i.e., the client may choose Eq.~\ref{eq:rule_extending_bitstring_size_y1} to encode data if the majority of $\boldsymbol{y}$ is $1$ to decrease the probability that $1$s are being detected.
Nevertheless, the trade-off for employing this redundant encoding approach manifests as an augmented quantum communication complexity, which reads $O(\frac{\log (NM)}{\epsilon})$.

\section{Construction of oracle operator $\hat{U}_{X_2}$ for blind QBC with untrusted client}\label{app: U_X_2}

In this section, we give the details for the implementation of $\hat{U}_{X_2}$ operator mentioned in Sec.~\ref{sec: untrusted_client}. 
Recall that $\hat{U}_{X_2}$ is applied to extract the phase term $(-1)^{x_i y_i}$, as shown in Eq.~\ref{eq:U_X_2}. 
The quantum state before applying $\hat{U}_{X_2}$ is given by 
\begin{equation}
    \frac{1}{\sqrt{N}}\sum_{i}^N \ket{i}_n (a_i\ket{0}+b_i (-1)^{y_i}\ket{1})_{o_1},
\end{equation}
 where $a_i$ and $b_i$ depends on $x_i$ and the encoding basis $R_i$. 
For the data $j$ encoded in Z basis, i.e., $R_j = 0$, one has $a_j b_j =0$ and the phase $(-1)^{x_i y_i}$ naturally shows up as in the original QBC algorithm. 
While for the data encoded in X basis, i.e., $R_j = 1$, we target to extract the $(-1)^{x_i y_i}$ term by transforming it back to Z basis. 

\begin{figure}[t]
\centering
\includegraphics[width=0.45\textwidth]{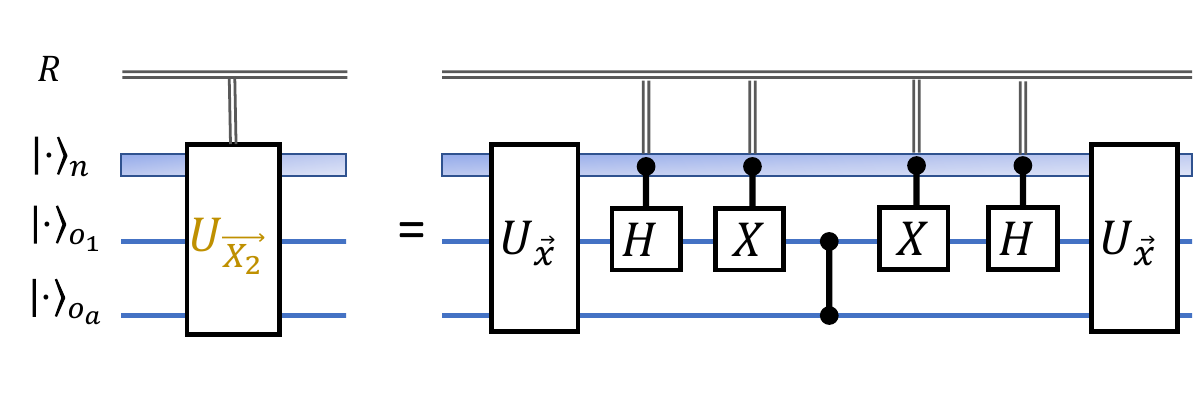}
\caption{Circuit diagram for implementing $U_{\Vec{X_2}}$ with the help of an ancilla qubit $o_a$. The first control line shows the classical control decided by the random number $R_i, i = 1, ... N$.  \label{fig: U_X2} }
\end{figure}

For this purpose, we consider the following protocol.
Firstly, $\hat{U}_{\Vec{x}}$ oracle is called to generate the state $\frac{1}{\sqrt{N}}\sum_{i}^N \ket{i}_n (a_i\ket{0}+b_i (-1)^{y_i}\ket{1})_{o_1} \ket{x_i}_{o_a}$ where the additional qubit $o_a$ encodes $x_i$ in Z basis. Secondly,  a Hadamard gate is applied on qubit $o_1$ conditioned on index qubit state $\ket{i}_n = \ket{j}_n$ that satisfies $R_{j}=1$ (i.e., encoding in X basis).  This will transform the X basis encoding to Z basis. Then, a NOT gate on qubit $o_1$ conditioned on those index qubit states  followed by a controlled-Z gate between $o_1$ and $o_a$ is applied. With the above steps, a phase $(-1)$ is generated unless $x_i=y_i=1$. The state now reads:
\begin{equation}
    \frac{1}{\sqrt{N}}\sum_{i}^N (-1)^{x_i y_i}\ket{i}_n \ket{m_i}_{o_1} \ket{{x_i}}_{o_a}.
\end{equation}
 Here $m_i=x_i$ when $R_i=0$ or when $R_i=1$ and $y_i=0$. 

Now that as the phase term $(-1)^{x_i y_i}$ has already been extracted, we transform the $o_1$ qubit state $\ket{m_i}_{o_1}$ back to the initial $(a_i\ket{0}+b_i (-1)^{y_i}\ket{1})_{o_1}$ by applying the controlled Hardmard and NOT gate again, and then decouple the ancillary qubit by calling the $\hat{U}_{\Vec{x}}$. The resulting quantum state reads $\frac{1}{\sqrt{N}} \sum_{i}^N (-1)^{x_i y_i}\ket{i}_n (a_i\ket{0}+b_i (-1)^{y_i}\ket{1})_{o_1}$.


\bibliography{blindQML_arXiv} 


\end{document}